\documentclass[aps,pra,superscriptaddress,superscriptaddress,preprint]{revtex4-1}

\usepackage{graphicx}% Include figure files
\usepackage{dcolumn}% Align table columns on decimal point
\usepackage{bm}% bold math
\usepackage{multirow}
\usepackage{upgreek}
\usepackage{braket}
\usepackage{bm}        % for math
\usepackage{amssymb}   % for math
\usepackage{amsmath}
\usepackage{subfigure}
\usepackage{epstopdf}
\usepackage{comment}
\usepackage{float}

\usepackage{xcolor} % Mark Edwards added color package!

\usepackage{hyperref}
\hypersetup{
    colorlinks=true,       % false: boxed links; true: colored links
    linkcolor=blue,          % color of internal links (change box color with linkbordercolor)
    citecolor=blue,        % color of links to bibliography
    filecolor=blue,      % color of file links
    urlcolor=blue           % color of external links
}

\newsavebox{\graphicsbox}

\usepackage{txfonts}

\usepackage{graphicx}% Include FIG. files
\usepackage{dcolumn}% Align table columns on decimal point
\usepackage{bm}% bold math

\begin{document}

%\preprint{Submitted to {\em New Journal of Physics}}

\title{Resonant wavepackets and shock waves in an atomtronic SQUID}

\author{Yi-Hsieh Wang$^{1,2}$, A.\ Kumar$^1$, F.\ Jendrzejewski$^{1,3}$, Ryan M.\ Wilson$^{1,4}$, Mark Edwards$^{1,5}$, S.\ Eckel$^1$, G.\ K.\ Campbell$^1$, Charles W.\ Clark}

\affiliation {Joint Quantum Institute, National Institute of Standards and Technology
and the University of Maryland, College Park, Maryland 20742,
\\$^2$Chemical Physics Program, University of Maryland, College Park, Maryland 20742,
\\$^3$Kirchhoff Institut f\"{u}r Physik, Ruprecht-Karls-Universit\"{a}t Heidelberg, 69120 Heidelberg, Germany,
 \\$^4$Department of Physics, The United States Naval Academy, Annapolis, MD 21402,
 \\$^5$Physics Department, Georgia Southern University, Statesboro, GA 30460}

\begin{abstract} The fundamental dynamics of ultracold atomtronic devices are reflected in their phonon modes of excitation. We probe such a spectrum by applying a harmonically driven potential barrier to a $^{23}$Na Bose-Einstein condensate in a ring-shaped trap. This perturbation excites phonon wavepackets. When excited resonantly, these wavepackets display a regular periodic structure. The resonant frequencies depend upon the particular configuration of the barrier, but are commensurate with the orbital frequency of a Bogoliubov sound wave traveling around the ring.  
Energy transfer to the condensate over many cycles of the periodic wavepacket motion causes enhanced atom loss from the trap at resonant frequencies.   Solutions of the time-dependent Gross-Pitaevskii equation exhibit quantitative agreement with the experimental data. We also observe the generation of supersonic shock waves under conditions of strong excitation, and collisions of two shock wavepackets.   
 
\end{abstract}

\pacs{Valid PACS appear here}% PACS, the Physics and Astronomy
\maketitle

%\tableofcontents

\section{Introduction}

In the emerging field of atomtronics~\cite{Seaman2007}, devices have now been realized that have counterparts in the realm of superconductivity~\cite{Jaklevic1964,Jaklevic1965,Silver1967}.  In particular, atomtronic devices based on ring-shaped Bose-Einstein condensates (BECs) with one or more weak links have been demonstrated.  Devices with one rotating weak link resemble the radio frequency superconducting quantum interference device (rf-SQUID), and exhibit similar physical effects like quantized persistent currents, phase slips and hysteresis.~\cite{Ramanathan2011,Wright2013,Eckel2014,Eckel2014b}.  Devices with two weak links more closely resemble the dc-SQUID, showing behavior consistent with the dc- and ac-Josephson effects and exhibiting clear signs of resistive flow above the critical mass current~\cite{Ryu2014,Jendrzejewski2014}.  However, the characteristic interference signal of the dc-SQUID has yet to be observed in an atomtronic device.
Although they use the same ring condensates and weak links, the one- and two-link devices display different critical velocities: that of the one-link system disagrees with 
the standard Bogoliubov speed of sound in a BEC, but that of the two-link system agrees with it~\cite{Eckel2014,
Jendrzejewski2014}.  

For BECs modeled with the zero-temperature Gross-Pitaevksii equation 
(GPE), the critical velocity is usually set by the bulk speed of sound associated with 
the low-energy phonon excitations of the condensate~\cite{Piazza2013,Eckel2014}.  In a ring without a weak link, surface modes of condensate
excitations can play an important role in setting the critical velocity~\cite{PhysRevA.86.011602}.
Other effects, such as the presence of a noncondensate fraction at finite temperature~\cite{Mathey2014} or condensate phase fluctuations~\cite{Mathey2010}, 
may also play a role in the dynamics of ring condensates.  For all of these outstanding issues, 
understanding the condensate excitation spectrum is crucial.  Such understanding may also lead to new devices. For example, 
a phonon interferometer was recently studied experimentally in Ref.~\cite{Marti2015}.
Additionally, knowledge of phonon spectroscopy may be useful in characterizing 
device performance, for example, in determining the circulation state of a ring BEC  \cite{Kumar2015}.

Here we report studies of excitations in a ring BEC driven harmonically by a potential barrier localized in a small region of the ring. This barrier is a variant of the weak-link structures that we have used in previous studies \cite{Jendrzejewski2014,Eckel2014,Eckel2014b}.  We find a number of resonant frequencies at which the driven condensate exhibits recurrent wavepacket trains traveling at the speed of sound. The resonant frequencies are multiples of the orbital frequencies of the wavepackets, which are easily calculated from the speed of sound and the symmetry of the driving potential. We construct a simple model of the resonant wavepackets that seems to have wide applicability. The resonant wavepackets persist over many cycles of the driven oscillation.  For sufficiently long excitation times, atoms eventually acquire enough energy to escape from the trap. We find that atom loss from the trap is strongly enhanced at the resonant frequencies, and that it is well described by solutions of the time-dependent GPE.  This suggests that our atom loss is dominated by the effects of mode-coupling that are known to exist in the strongly driven GPE \cite{PhysRevLett.78.3589,Choi1998,Morgan98,proukakis2013quantum}.  For strong conditions of excitation, we observe wavepackets that move faster than the speed of sound.  These resemble the shock waves that have been seen in previous studies \cite{PhysRevLett.101.170404,PhysRevA.80.043606}. In this system we also observe the collision of two shock waves.

Sec.~\ref{sec:simple_model} presents an intuitive picture of the dynamics of harmonically driven ring BECs in terms of phonon wavepackets.  Sec.~\ref{sec:method} describes the details of our experimental setup.  In Sec.~\ref{sec:model}, we model the condensate dynamics with the GPE and the condensate's elementary excitations with the Bogoliubov-de Gennes equations.  Sec.~\ref{sec:driving_excitations} shows that appropriate modulations of the potential barrier can be used for controlled  excitation of resonant wavepackets.  We present evidence for the generation of supersonic shock waves and collisions of two shock wavepackets in Sec.~\ref{sec:shock_waves}.

\section{\label{sec:simple_model} A simple model of resonant wavepacket generation by an oscillating weak link}

In a Bose-Einstein condensate, phonon wavepackets can be constructed by forming a superposition of low-lying Bogoliubov 
excitations (see Sec.~\ref{sec:model} for details).  If a localized weak perturbation with a typical length scale $w$ is suddenly applied to the condensate ~\cite{Andrews1997}, it will generate a  
wavepacket consisting of phonon modes with wavelengths less than $w$.  These wavepackets will travel away from the perturbation at the speed of sound without dispersion. Stronger perturbations can generate a variety of nonlinear wave motions \cite{DNSE2015}, including supersonic shock waves that are discussed in Sec.~\ref{sec:shock_waves}.

In a ring condensate, one can create wavepackets that travel around the ring.    
The orbital period, $T$, of a single wavepacket establishes a characteristic frequency, $\nu = 1/T = c/(2\pi R)$, where $c$ is the speed of sound in the BEC and $R$ is the radius of the ring. 
If the localized weak perturbation is modulated periodically, resonances may occur when the perturbation consistently adds energy to the wavepacket over each cycle of its motion, as is the case in cyclotron and synchrotron particle accelerators \cite{Wiedemann2015}.  Two classes of such resonances, associated with amplitude- and position-modulation of a weak link,  have been seen in our experiment. This section presents a simple model for understanding them.

\begin{figure}
\centering
\includegraphics[width=5in]{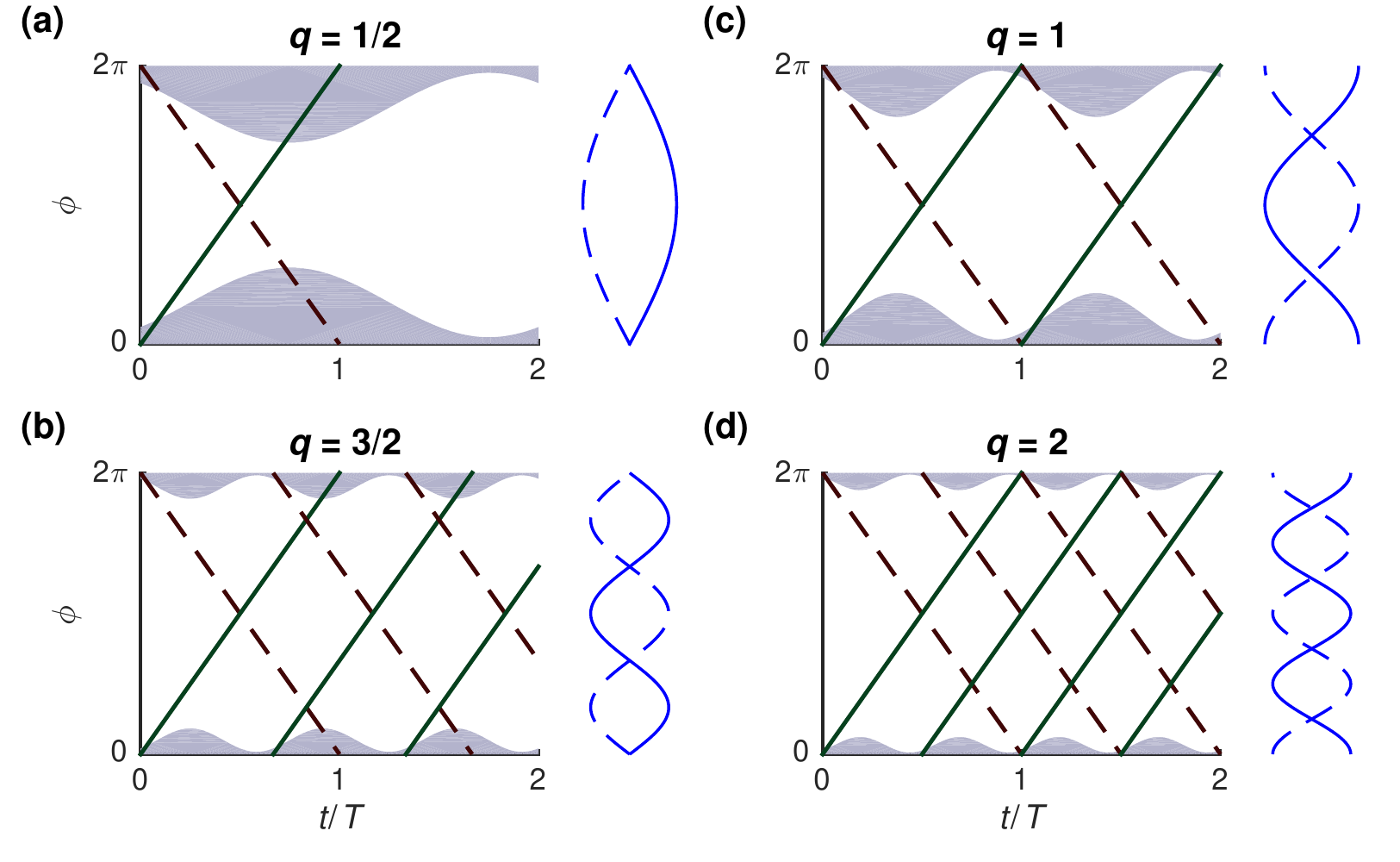}\\ 
\caption{\label{fig:wavepacket_oscim} Trajectories of wavepackets in a ring driven by a barrier whose height oscillates with frequency $\nu_q = q\nu$: (a) $q=1/2$; (b) $q=3/2$; (c) $q=1$; (d) $q=2$. The grey shaded areas represent the barrier height as a function of time, $t$, and the solid (green) and dashed (black) lines indicate the ring azimuthal coordinates, $\phi$, of the centers of the wavepackets.   The blue lines show the standing-wave-like density modulation created by the overlapping wavepackets.
}
\end{figure}

Consider first the case where the perturbation is a symmetric potential barrier  at a fixed location in the ring, with the barrier height driven sinusoidally around a positive value.  This amplitude-modulation case is shown schematically in Fig.~\ref{fig:wavepacket_oscim}.  Here, the widths of the gray-shaded regions at $\phi = 0$ and $2\pi$ denote the height of the barrier as a function of time.  As the height increases, the barrier displaces the BEC, generating a symmetric pair of wavepackets.  Traveling in opposite directions around the ring, the wavepackets return to the barrier at time $T$.  If the barrier is rising when they return, energy will be added to the wavepackets as they begin their next journey around the ring.  The resonance condition in this case corresponds to the frequency of the barrier oscillation being an integer multiple of $\nu$, i.e., $\nu_q = q \nu$, where $q$ is an integer.  Panels (c) and (d) of Fig.~\ref{fig:wavepacket_oscim} depict this resonant case with $q=1$ and $2$, respectively.  On the other hand, if the barrier is falling when the wavepackets arrive, they will lose energy. This non-resonant case corresponds to  $\nu_q = q \nu$, where $q$ is now a half-integer.  Panels (a) and (b) of Fig.~\ref{fig:wavepacket_oscim} show this non-resonant case with $q=1/2$ and $3/2$, respectively. 

When the counterpropagating wavepackets overlap, they create a localized region of high density.  The alternation of regions of high and low densities follows a pattern analogous to a standing wave, which is shown by the dashed and solid blue curves to the right of each panel. For panels (c) and (d), this standing density wave propagates like $\cos(q \phi)\sin(2 \pi  \nu_q t)$.  Since their wavefunctions resemble the eigenfunctions of a particle on a ring, we denote these modes as `ring modes'.  
For the nonresonant conditions shown in panels (a) and (b), the density wave  propagates like $\sin(q\phi)\sin(2 \pi \nu_q t)$.  Given their similarity to the eigenfunctions of a particle in a box potential, we denote them as `box modes'. For amplitude-modulation excitation, the ring modes are resonant and the box modes are nonresonant.

\begin{figure}
\centering
\includegraphics[width=5in]{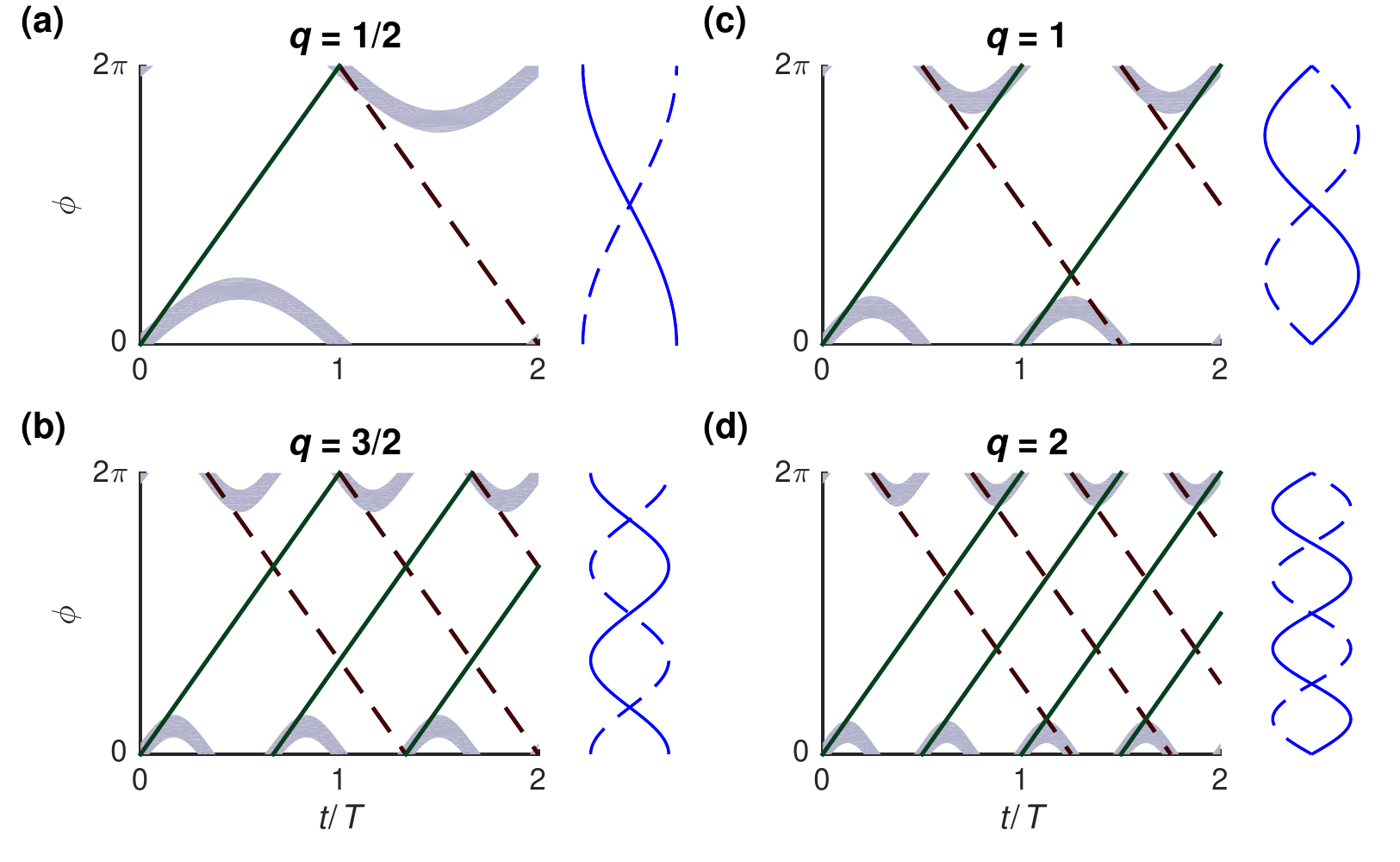}\\ 

\caption{\label{fig:wavepacket_oscip} Trajectories of wavepackets in a ring driven by a barrier whose position oscillates with frequency $\nu_q = q\nu$: (a) $q=1/2$; (b) $q=3/2$; (c) $q=1$; (d) $q=2$.   The grey shaded areas represent the barrier position as a function of time, $t$, and the solid (green) and dashed (black) lines indicate the azimuthal coordinate, $\phi$, of the centers of the wavepackets.  The blue lines show the standing-wave-like density modulation created by the overlapping wavepackets.
}
\end{figure}

We now consider position-modulation excitation, in which the shape of the barrier remains constant, but its azimuthal position in the ring oscillates in time.
In this case, a wavepacket is created by the barrier pushing atoms in front of it during the fastest part of its motion.  The wavepacket then orbits the ring with period $T$.  On its return, if the wavepacket encounters the barrier moving opposite to its direction of travel, it will be reflected from the barrier with a momentum kick. 
Therefore, the resonance condition in the position-modulation case corresponds to the frequency of wavepacket oscillation being a half-integer multiple of $\nu$, i.e. $\nu_q =q \nu$ with $q=1/2$, $3/2$, etc.  Fig.~\ref{fig:wavepacket_oscip} shows the wavepacket propagation for position-modulation, in a manner similar to Fig. \ref{fig:wavepacket_oscim}.  Panels (a-d) correspond to $q= 1/2$, 1, 3/2, and 2, respectively.  The corresponding standing waves are shown on the right of each panel. For position-modulation excitation,  the box modes are resonant and the ring modes are nonresonant.

The above argument for the resonance condition of a position-modulated barrier is valid only if the wavepacket is predominantly reflected from the barrier.  If, on the other hand, the wavepacket were predominantly transmitted, the barrier would only add energy to the wavepacket if it were traveling in the same direction as the wavepacket.  In this case, the resonance condition would again correspond to the oscillation frequency being an integer multiple of $\nu$.  (In this case, the ring modes would be again resonant.)  Whether the wavepacket is transmitted or reflected depends on the change in the speed of sound in the barrier region.  If the barrier is strong compared to the chemical potential, the density will be depleted, and the speed of sound would be reduced.  In analogy to optics, the change in the index of refraction going from the ring to the barrier would be large, consequently causing a large reflectivity.  If the barrier is weak compared to the chemical potential, the index of refraction change would be small, minimizing the amount of reflection.  Thus, one should expect a change in the resonance condition for a position-modulated barrier: as the strength of the barrier potential is increased, the resonance should shift from integer to half-integer values of $q$.

\section{\label{sec:method} Experimental Parameters}

The BEC is formed in a crossed optical dipole trap with the same procedures as in Ref.~\cite{Eckel2014b}. The trap is created by two laser beams: a red-detuned laser beam shaped like a sheet for the vertical confinement and a blue-detuned laser beam transmitting through an intensity mask~\cite{Lee2014}.  The intensity mask is imaged onto the atoms, providing in-plane confinement.  Laser-cooled $^{23}$Na atoms in the $|F=1,M_F=-1\rangle$ state are condensed into the trap after forced evaporation.

The intensity mask forms a ``target''-shaped trap~\cite{Eckel2014b}, which has both a toroidal (ring) and a concentric disc trap.  The resulting condensate in the ring has a mean radius of 22(1)~$\mu$m and a Thomas-Fermi full-width of $\approx8$~$\mu$m. The disc-shaped condensate, which is left unperturbed during the experiment, serves as a reference to check atom number stability. The vertical trapping frequency is $\omega_z/2\pi = 542 (13)$~Hz while the radial trapping frequency of the ring is $\omega_\rho/2\pi = 400 (20)$~Hz.  The average number of atoms in the target trap is $\approx7 \times 10^5$, with $\approx80$\% of atoms in the toroid and $\approx20$\% in the central disk.  On any individual repetition of the experiment, the atom number can fluctuate from its mean by up to $10$\% ($2\sigma$). 
We measure the atomic density using {\it in-situ} partial transfer absorption imaging~\cite{Ramanathan2012}. 

We create the weak link potential by using a focused, blue-detuned Gaussian beam.  The $1/e^2$ full-width of the Gaussian is $\approx5$~$\mu$m.  This beam generates a repulsive potential that depletes the condensate density locally in the region of the beam.   An acousto-optic deflector (AOD) controls the position of the beam.  By changing the power applied to the AOD, we can control the intensity of the beam.  To create a radially-elongated weak link, the AOD scans the beam rapidly in the radial direction at 2~kHz.  The resulting time-averaged potential is a wide, flat potential barrier with an effective width of $\approx15$~$\mu$m.

For the experiments here, we manipulate the weak link in a variety of different ways.    For the experiments described in Sec.\ref{sec:driving_excitations}, the weak link is first applied adiabatically to the BEC, so as to not generate excitations.  The weak link beam's intensity is ramped on linearly  over 300~ms.  During this linear ramp, the azimuthal position of the weak link is fixed.  After the intensity reaches its final value, the weak link position is oscillated in the azimuthal direction, or its intensity is modulated as a function of time.  For the experiments described in Sec.~\ref{sec:shock_waves}, the weak link beam is turned on suddenly while it remains in a fixed azimuthal position.  The response time of the AOD and the servo that controls the intensity of the beam limits the rise time of the weak link to approximately 100~$\mu$s.

\section{\label{sec:model} Bogoliubov - de Gennes desciption of elementary excitations of a BEC\\}

Dilute Bose-Einstein condensates, like the one we study here, are often \cite{stringari,pethick} described by a mean-field theory with an order parameter $\Psi_0 = \langle \hat{\Psi} \rangle$, where $\hat{\Psi}$ is the quantum field operator, which can be expanded as
\begin{eqnarray}
\label{eq:BdG0}
\hat{\Psi} &=&\Psi_{0}+ \delta\hat{\Psi}.
\end{eqnarray}
Here, $\delta\hat{\Psi}$ denotes the field operator for the non-condensate atoms: it
describes the elementary excitations of the condensate in the linear-response regime.  The order parameter $\Psi_0$ can be interpreted physically as the condensate wave function, which is macroscopically occupied.  The dynamics of the condensate wave function are described by the   time-dependent Gross-Pitaevskii equation (TDGPE),
\begin{eqnarray}
\label{eq:GP}
i\hbar\frac{\partial }{\partial t}\Psi_0(\mathbf{r},t)&=&\left [ -\frac{\hbar^2}{2M}\nabla^2+V(\mathbf{r})+g\left | \Psi_0(\mathbf{r}) \right |^2\right ]\Psi_0(\mathbf{r}),
\end{eqnarray}
where $V\left(\mathbf{r}\right)$ is the external potential, $M$ is the atomic mass, and $g$ quantifies the interaction strength between the atoms, and is given by $g=4\pi\hbar^2 a/M$, where $a$ is the $s$-wave scattering length associated with binary atomic collisions. The stationary, ground state solution of the GPE can be expressed as
\begin{eqnarray}
\label{eq:psi}
\Psi_0(\mathbf{r},t)=\sqrt{n(\mathbf{r})} e^{-i\mu t/\hbar}
\end{eqnarray}
where $n(\mathbf{r})$ is the condensate density, and $\mu$ is the chemical potential of the system.

Elementary excitations are those for which the number of excited atoms is much smaller than the number of atoms in the BEC.  The field operator for the excited atoms then satisfies the linearized equation of motion,
\begin{eqnarray}
\label{eq:fluctuation}
i\hbar\frac{\partial }{\partial t}\delta\hat{\Psi}(\mathbf{r},t)&=&\left [ -\frac{\hbar^2}{2M}\nabla^2+V(\mathbf{r})+2gn(\mathbf{r})-\mu \right ]\delta\hat{\Psi}(\mathbf{r},t)+ g \Psi_0^2 \delta\hat{\Psi}^{\dagger}.
\end{eqnarray}
We solve this equation in a Bogoliubov-de Gennes framework \cite{stringari,pethick}, and expand  $\delta\hat{\Psi}$ in the Bogoliubov operators $a_i$ and $a^\dagger_i$,
\begin{eqnarray}
\label{eq:BdG1}
\delta\hat{\Psi} &=& \sum_{i} \left( u_{i} e^{-i\omega_{i} t} a_{i} +v_{i}^{*} e^{i\omega_{i} t}a_{i}^{\dagger}\right),
\end{eqnarray} 
where $\omega_{i}$ are the elementary excitation frequencies and $u_i$, $v_i$ are the Bogoliubov amplitudes of the $i^\mathrm{th}$ excitation.
These amplitudes satisfy the Bogoliubov-de Gennes (BdG) equations:
\begin{eqnarray}
\label{eq:BdG2}
\left [\hbar\omega_i+\frac{\hbar^2}{2M}\nabla^2-V(\mathbf{r})-2gn(\mathbf{r})+\mu \right ]u_i(\mathbf{r})&=&gn(\mathbf{r})v_i(\mathbf{r})\nonumber\\ 
\left [-\hbar\omega_i+\frac{\hbar^2}{2M}\nabla^2-V(\mathbf{r})-2gn(\mathbf{r})+\mu \right ]v_i(\mathbf{r})&=&gn(\mathbf{r})u_i(\mathbf{r}).
\end{eqnarray} 
We numerically diagonalize Eqs.~(\ref{eq:BdG2}) to find the spectrum of elementary excitations for the condensate.

In a spatially uniform condensate ($V(\mathbf{r})=0$), Eq.~(\ref{eq:BdG2}) has plane-wave solutions~\cite{pethick}, $u_{\mathbf{k}}({\mathbf{r}})=u_{\mathbf{k}} e^{i\mathbf{k}\cdot \mathbf{r}}$ and $v_{\mathbf{k}}({\mathbf{r}})=v_{\mathbf{k}} e^{i\mathbf{k}\cdot\mathbf{r}}$, where $\mathbf{k}$ denotes the wave vector, and a continuous spectrum of elementary excitations,
\begin{eqnarray}
\label{eq:dispersion}
\hbar\omega_{\mathbf{k}} &=& \sqrt{\epsilon_{\mathbf{k}}^2+2\epsilon_{\mathbf{k}} gn},
\end{eqnarray}
where $\epsilon_{\mathbf{k}}=\hbar^2 k^2/2M$ is the kinetic energy of a free quantum particle of mass $M$. For small $k$, the frequencies $\omega_{\mathbf{k}}$ are linear in $k$, i.e. $\omega_{\mathbf{k}} \approx k \sqrt{gn/M} $. Since this dispersion relation is the same as that for a sound wave, $\omega=ck$, these excitations, or quasiparticles, can be viewed as phonons, and the proportionality constant determines the speed of sound, $c=\sqrt{gn/M}$. 

In our experiments, we do not deal with a homogeneous BEC, but one confined to a ring trap.  The trap's potential has the form
\begin{eqnarray}
\label{eq:trap}
V(\mathbf{r})=\frac{1}{2}M\omega_z^2 z^2+V_G\left(1-e^{-2(\rho-R)^2/w_\rho^2}\right),
\end{eqnarray}
where the first term is a harmonic potential in the axial ($z$) direction with frequency $\nu_z =  542$~Hz, and the second term represents the ring potential, with: depth $V_G=266 \mathrm{nK} \times k$, where $k$ is the Boltzmann constant; ring radius $R=22.4$~$\mu$m; and $1/e^2$ half-width $w_\rho=5.5$~$\mu$m.  (These parameter values best reproduce the experiment as described in Sec.~\ref{sec:method}.)   In this trap, the low-momentum quasiparticles obey a {\em quantized} dispersion relation analogous to that of the homogeneous system given by Eq.~\ref{eq:dispersion}.  We now describe this correspondence.  
  
In the absence of a weak link, the potential and the ground state of the BEC have cylindrical symmetry about the $z$ axis. The Bogoliubov quasi-particle amplitudes thus have sharp values of the projection of the angular momentum operator $\hat{l}_z = \hat{x}\hat{p}_y - \hat{y}\hat{p}_x$, and thus have the azimuthal dependence of the form $\sim e^{i m \phi}$, where $\phi = \arctan(y/x)$ is the conventional azimuthal angle of a two-dimensional coordinate system and $m$ is an integer.  Our lowest energy solutions to Eqs.~(\ref{eq:BdG2}) scale like $\omega_m \sim m \sqrt{g \bar{n} / M R^2}$, 
where $\bar{n}$ is the mean condensate density and $R$ is the radius of the ring \cite{Zaremba1998}.  These solutions form a manifold of discrete phonon-like modes that propagate azimuthally with the characteristic speed of sound of the ring condensate, 
\begin{equation}
\label{eq:speedosound}
c = \sqrt{g \bar{n} / M}.
\end{equation}  
  
\begin{figure}[t]
\centering
\includegraphics[width=4in]{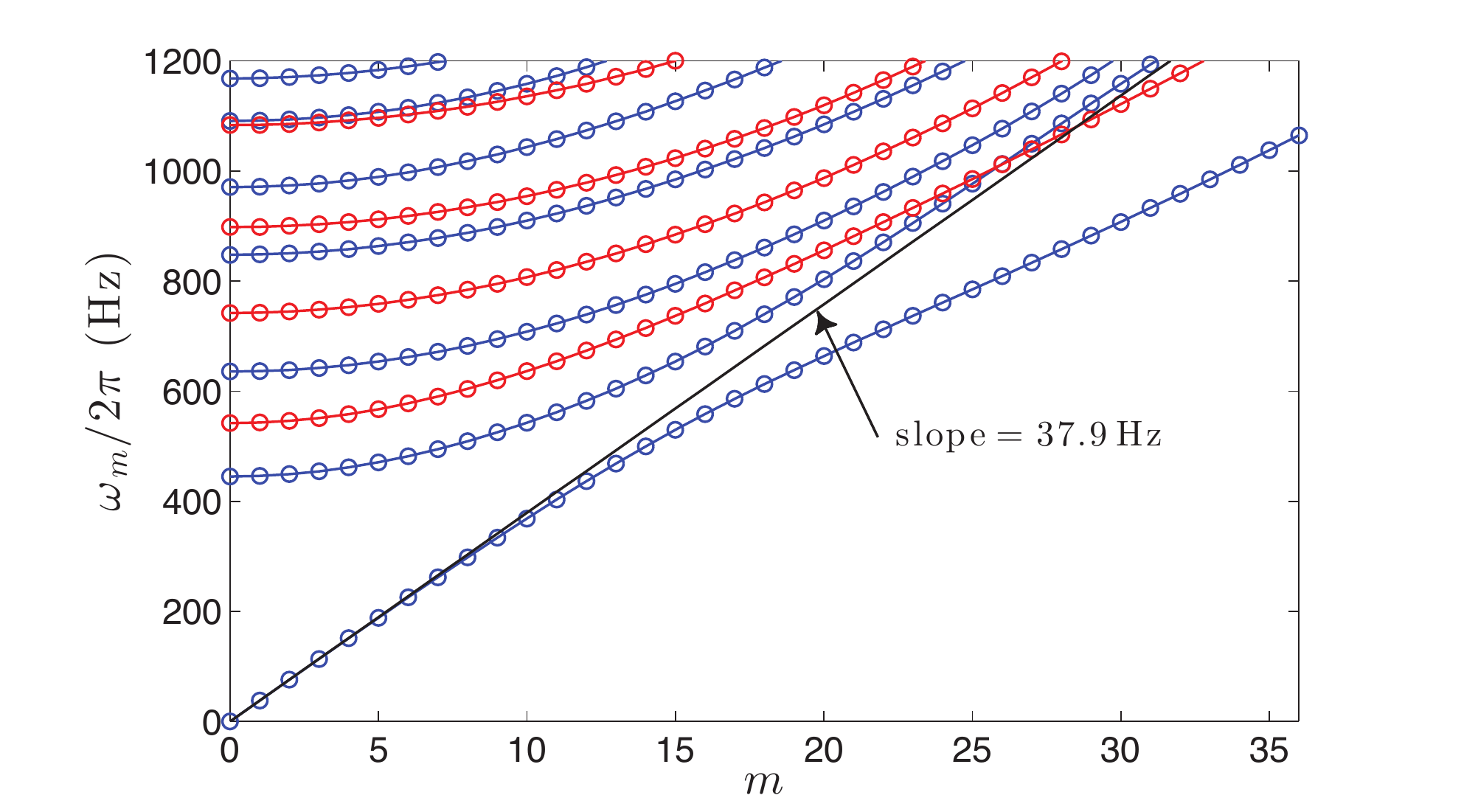}\\% Here is how to import EPS art
\caption{\label{fig:BdG_spec} The BdG spectrum for the elementary excitations of a ring condensate. The blue (red) curves correspond to the excitation modes that are even (odd) in the axial ($z$) direction. The lowest branch represents the excitations in the azimuthal direction, the frequency of which is linear at small $m$ (denoted by the black line). The slope determines the orbital frequency of sound $\nu=37.9(2)$~Hz.}
\end{figure}

Fig.~\ref{fig:BdG_spec} shows the calculated the energy spectrum of elementary excitations by solving the BdG equations.  For small $m$, the modes for the lowest branch are nodeless in the radial and axial ($z$) directions \cite{PhysRevA.86.011602}.  The linear dependence at small $m$ is clear.  Using the experimental parameters of Sec.~\ref{sec:method}, our linear fit at small $m$ provides an orbital frequency for sound  of $\nu = 37.9(2)$~Hz. Note the avoided crossing between the lowest two branches around $m = 17$, which is due to the near degeneracy of angular and radial excitation there. In a manner characteristic of two-level crossing systems \cite{Clark1979}, the predominantly angular modes continue on the second branch for $m > 20$, where they show roughly the same linear dispersion relation.

We theoretically model the procedure described in Sec.~\ref{sec:simple_model} by propagating Eq.~(\ref{eq:GP}) in imaginary time to find the ground state condensate wave function $\Psi_0(\mathbf{r})$, then propagating in real time to model the dynamics.  We implement the split-step Crank-Nicholson algorithm as in \cite{adhikari2009} on a Cartesian grid of dimensions $x \times y \times z =100\times 100\times 10$ $\mu \text{m}$.  To generate various wavepacket trajectories, we generalize to a time-dependent potential $V(\mathbf{r},t)$.  This potential includes both the static potential (Eq.~\ref{eq:trap}) and a potential for the  weak link.  This latter potential is repulsive and includes a Gaussian of $1/e^2$ half-width $w_L=5$~$\mu$m along the azimuthal direction, and a rectangle of width $L=15$~$\mu$m along the radial direction (see the supplemental material of Ref.~\cite{Eckel2014}).

During the simulation, atoms can gain sufficient energy to escape from the trap.  Such atoms are nonetheless bounded in the box, and can reflect at the boundaries and subsequently return to the ring trap.  To eliminate this numerical effect, we implement absorbing layers at the edges of the $xy$-plane of the gridded box, by adding a damping term $H_{damp}=i\Gamma(x,y)$ in the Hamiltonian.  We adopt the damping term that follows the form in \cite{kosloff1986}, in which $\Gamma(x,y)$ slowly increases as $x$ or $y$ approach the box boundaries.  In the $x$ direction (equivalently for $y$),  $\Gamma=V_d/\cosh^2(|x-x_d|/L_d)$, where the damping constant $V_d$ is taken to be 0.01 $\mu$, the absorbing layer width $L_d=10$~$\mu$m, and $|x-x_{\text{d}}|$ is the distance of a point in the absorbing layers from the nearest box boundary, $x_d$.

\section{\label{sec:driving_excitations} Driving and Probing the Excitations }

\begin{figure}
\centering
\includegraphics[width=6.5in]{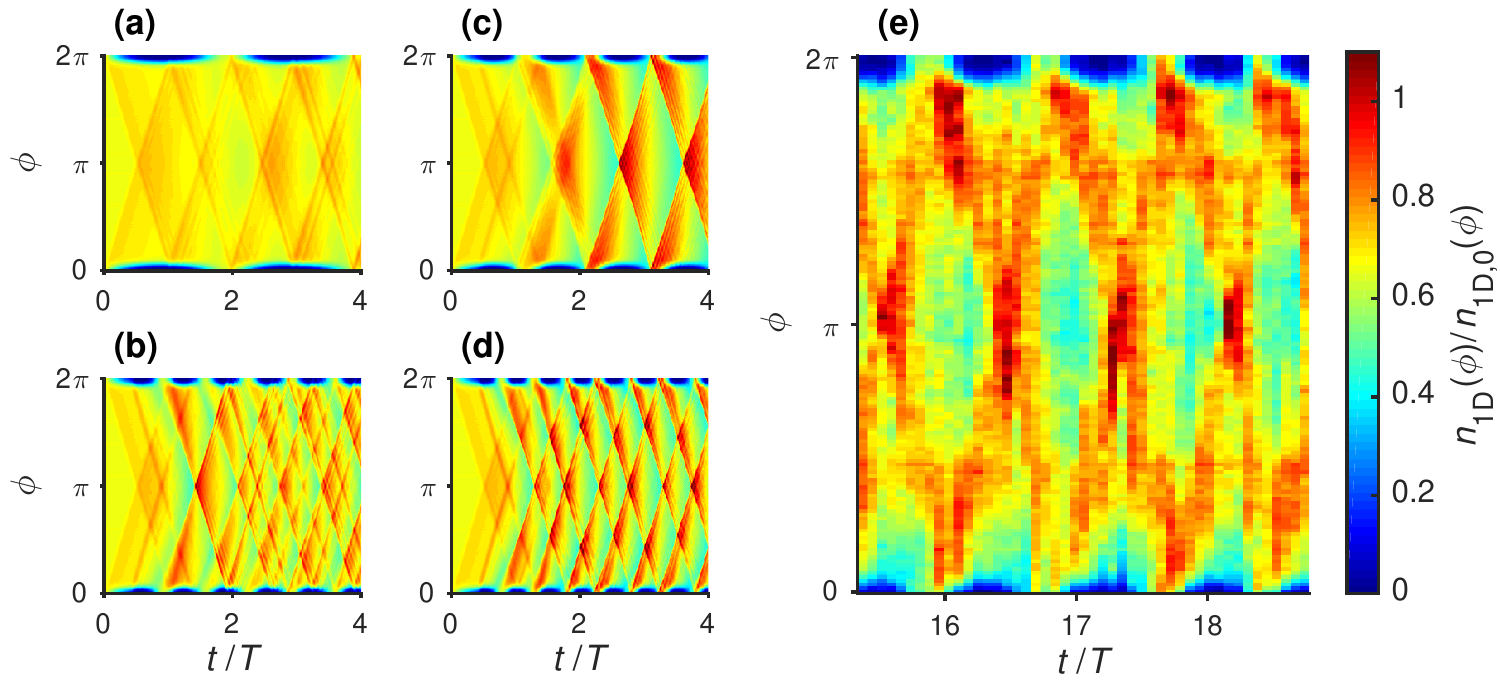}\\ 
\caption{\label{fig:oscim} Time evolution of wavepackets generated through amplitude modulation of the barrier (see Fig. \ref{fig:wavepacket_oscim}). The normalized 1D density  (colorbar) shows wavepackets, or localized regions of high density, moving around the ring (azimuthal coordinate $\phi$) with time $t$.  The density also shows the barrier oscillating at $\phi=0$ with frequency $\nu_q = q\nu$.  Modes with $q=1/2$ (a) and $q=3/2$ (b) are nonresonant; modes with $q=1$ (c and e) and $q=2$ (d) are resonant. }
\end{figure}

Knowing the orbital frequency $\nu=37.9$~Hz (Sec.~\ref{sec:model}), we now proceed to oscillate the barrier to find the resonant frequencies for wavepacket propagation (Sec.~\ref{sec:simple_model}).  This oscillation can take on two different forms. The first, shown schematically in Fig.~\ref{fig:wavepacket_oscim}, is the amplitude-modulation case discussed in Sec.~\ref{sec:simple_model}. The barrier height is given by  $V_b(t)=V_0+V_a\sin(2\pi v_q t)$, where $V_0/\mu=0.54(5)$ is the average amplitude of the barrier, $V_a=0.95 \, V_0$ is the amplitude of modulation, and $\nu_q=q\nu$ is the drive frequency.  (The uncertainty in $V_0$ applies only to the experimental value.) The barrier is ramped up to $V_0$ in 10 ms at the beginning of each evolution. Fig. \ref{fig:oscim} shows the resulting time evolution of the condensate density for both the experiment (e) and the GPE simulations (a)-(d).  At each time $t$, we integrate the condensate density along the radial and vertical directions to obtain an integrated 1D density $n_{1D}(\theta)$ along the azimuthal direction.  The normalized density shown in Fig.~\ref{fig:oscim} is then obtained by dividing $n_{1D}(\theta)$ by $n_{1D,0}(\theta)$, the 1D density measured in a unperturbed ring without a weak link.  In the experiment, the condensate density is not clearly periodic until several cycles of the oscillation have elapsed; therefore, we show data for later times $t\approx 15.5 \, T$ to $t\approx 19 \, T$.  At these later times, atoms have already left the trap causing the measured normalized densities to tend to be less than unity.  (For clarity, we scale the theoretical predictions to have the same range of normalized densities as the experiment.)  Because the wavepackets in the $q=1$ and $q=2$ cases collide with the accelerating barrier, each oscillation cycle increases the energy of the wavepacket.  Therefore, these ring modes are on resonance, as predicted in Sec.~\ref{sec:simple_model}.  The experimental data (Fig.~\ref{fig:oscim}e) shows the resonant $q=1$ mode, which is consistent with the GPE simulation (Fig.~\ref{fig:oscim}c).  

\begin{figure}
\centering
\includegraphics[width=6.5in]{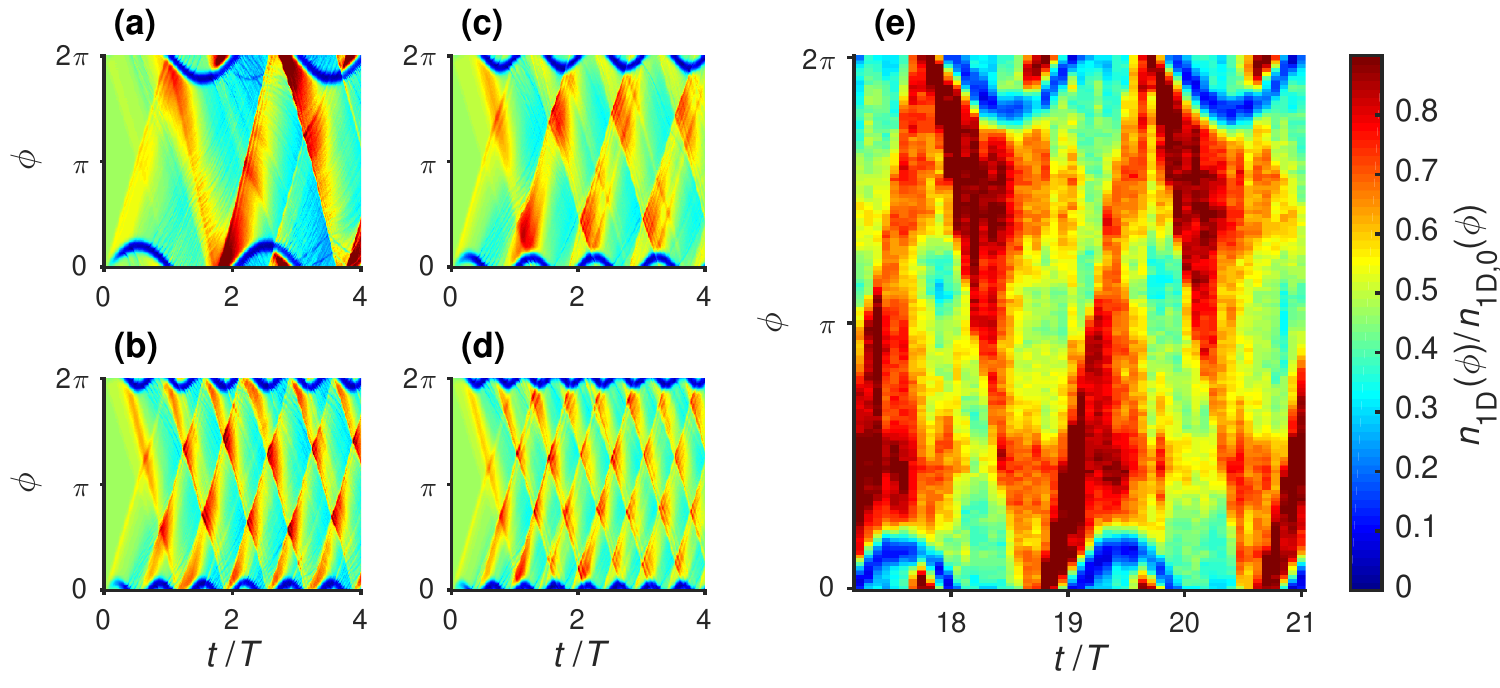}\\ 
\caption{\label{fig:oscip} Time evolution of wavepackets generated though position modulation of the barrier (see Fig. \ref{fig:wavepacket_oscip}).  The normalized 1D density (colorbar) shows wavepackets, or localized regions of high density, moving around the ring (azimuthal coordinate $\phi$) with time $t$.  The density also shows the barrier oscillating about $\phi=0$ with frequency $\nu_q = q\nu$.  Modes with $q=1/2$ (a and e) and $q=3/2$ (b) are resonant; modes with $q=1$ (c) and $q=2$ (d) are nonresonant.}
\end{figure}

The second oscillation scheme, shown schematically in Fig.~\ref{fig:wavepacket_oscip}, is the position-modulation case discussed in Sec. \ref{sec:simple_model}. Here, the position of the barrier is given by $\phi_b(t)=\phi_0+\phi_a\sin(2\pi \nu_q t)$, where $\phi_0=0$ is the average position of the maximum height of the barrier, $\phi_a$ is the amplitude of modulation, and $\nu_q = q \nu$ is the drive frequency.  The amplitude satisfies $\nu_q\phi_a=80$~rad/s, which ensures the maximum velocity of the barrier is independent of $\nu_q$.  Fig.~\ref{fig:oscip} shows the resulting  time evolution of the density for both the GPE simulations (a-d) and the experiment (e).  Here, the barrier, with height $V_0/\mu = 0.65(7)$, appears to be mostly reflective.  As predicted in Sec.~\ref{sec:simple_model}, the cases $q=1/2$ and $q=3/2$ (box modes) are on resonance.  In particular, the wavepacket trajectories are synchronized with the barrier motion: a wavepacket generated by the barrier propagates around the ring and collides with the barrier while the barrier is moving in the direction opposite the  wavepacket.  By  contrast, the cases $q=1$ and $q=2$ (ring modes) are off resonance: the wavepackets collide with the barrier at a point in its oscillation when it is moving in the same direction. 

Because the oscillating barrier can continually add energy to the condensate, atoms can acquire sufficient energy to escape the trap.  If the oscillation is resonant, efficient energy transfer from the barrier will result in atom loss.  Atom loss measured as a function of driving frequency will therefore show clear peaks at the resonant frequencies, $\nu_q$.  For both the experiment and the GPE simulation, we extract this atom-loss spectrum by oscillating the barrier for 2~s and then counting the remaining atoms in the trap, $N_R$.  The fraction of atoms 
that remain is given by $N_R/N$, where $N$ is the number of atoms measured when there is no oscillation. 

\begin{figure}
\centering
 \includegraphics[width=6in]{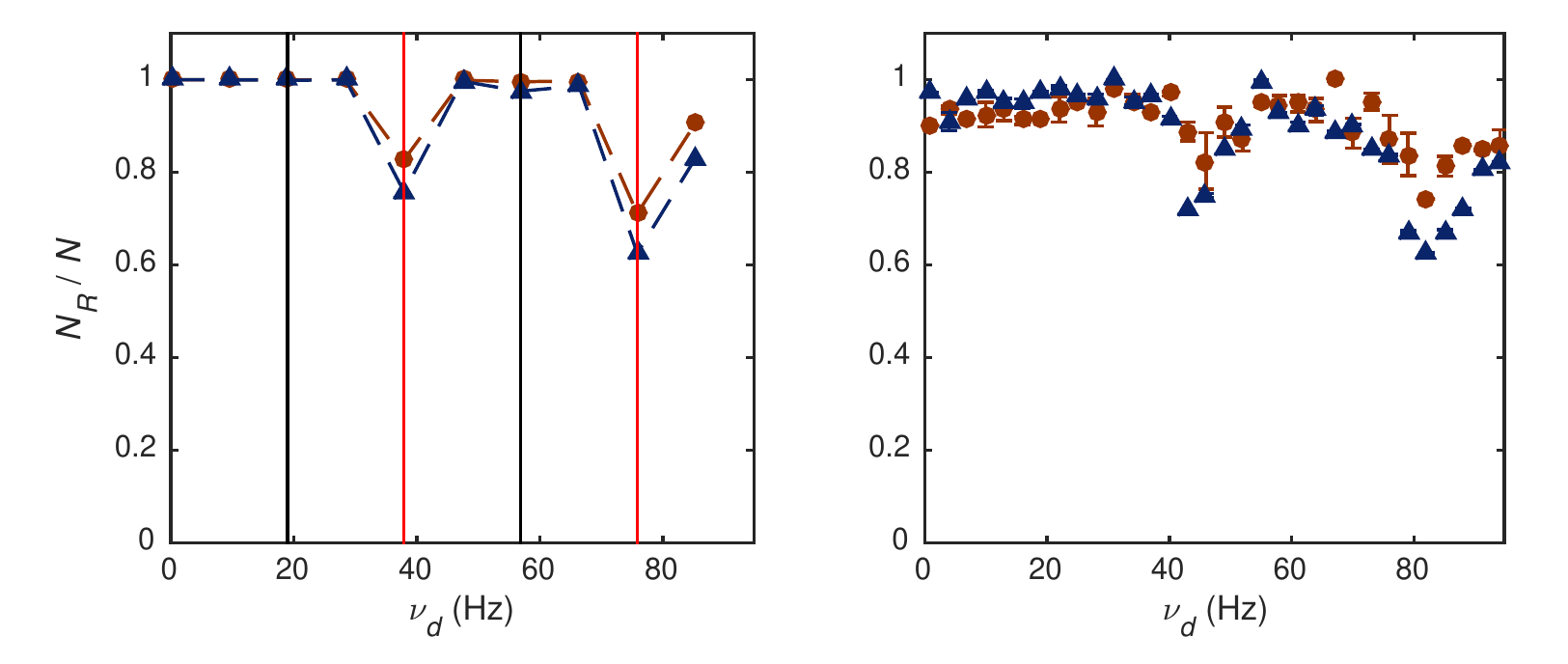}
\caption{\label{fig:loss_spec_height} Simulated (left) and experimental (right) atom-loss spectra for an amplitude-modulated barrier with frequency $\nu_d$ and average heights $V_0/\mu = 0.30(2)$ (blue triangles) and $V_0/\mu = 0.50(4)$ (red circles).  Here, $N_R/N$ is the fraction of atoms that remain in the trap after 2~s of excitation.  The vertical black (red) lines correspond to the resonant frequencies of the box (ring) modes.  The dashed lines are a guide to the eye.}
\end{figure}

\begin{figure}
\centering
 \includegraphics[width=6in]{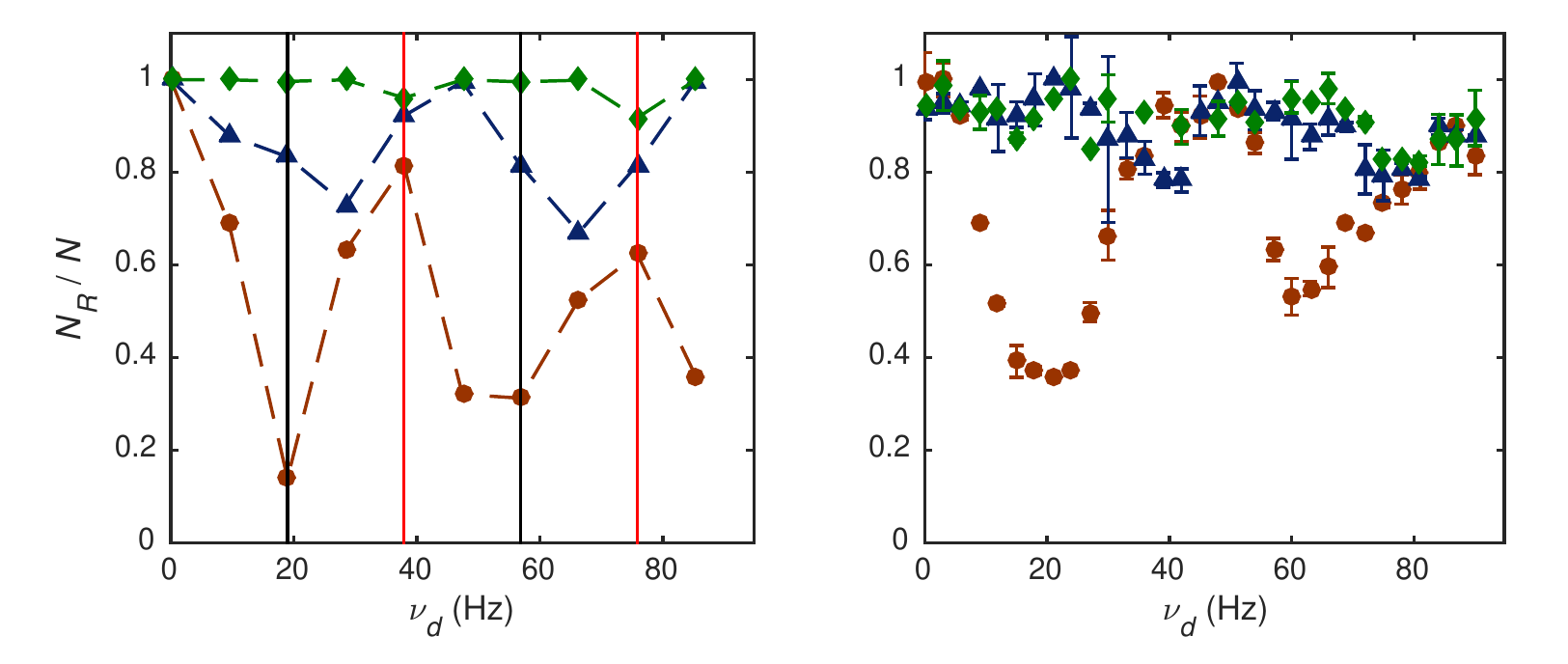}
\caption{\label{fig:loss_spec_position}
Simulated (left) and experimental (right) atom-loss spectra for a position-modulated barrier with frequency $\nu_d$ and heights $V_0/\mu=0.15(1)$ (green diamonds), $V_0/\mu = 0.30(2)$ (blue triangles), and $V_0/\mu = 0.60(4)$ (red circles).  Here, $N_R/N$ is the fraction of atoms that remain in the trap after 2~s of excitation.  The vertical black (red) lines correspond to the resonant frequencies of the box (ring) modes.  The dashed lines are a guide to the eye.} 
\end{figure}

Fig.~\ref{fig:loss_spec_height} shows the resulting atom-loss spectra obtained by amplitude modulation for two different barrier heights $V_0/\mu=0.30(2)$ and $0.50(4)$.  (Here, as before, the uncertainty applies only to the experiment.)  In both cases, the oscillation amplitude is given by $V_a=0.5V_0$. Both the experimental and simulated spectra show resonance peaks at drive frequencies corresponding to $q=1$ and $q=2$.  As expected, these are the resonant frequencies of the ring modes (Fig.~\ref{fig:wavepacket_oscim}c-d and Fig.~\ref{fig:oscim}c-d). 
The location of the peaks in the experiment indicates $\nu \approx 41$~Hz, slightly larger than that predicted by theory.  This small discrepancy may be due to uncertainty in atom number, trapping frequencies, or other experimental parameters.

Fig.~\ref{fig:loss_spec_position} shows the atom loss spectra for position-modulation with barrier heights $V_0/\mu=$ 0.15(1), 0.3(2), 0.6(4).  The displacement amplitude of the position modulation obeys $\nu\phi_a=60\text{ rad/s}$.  As $V_0$ is increased, both the experiment and the simulation show initial peaks at $q=1$ and $q=2$ that shift to $q=1/2$ and $q=3/2$.  This corresponds to a transition from the ring modes being resonant to the box modes being resonant.  For small barrier heights, wavepackets are predominantly transmitted through the barrier.  In this case, the ring modes are resonant, as these wavepackets receive more energy when they collide with a co-moving barrier (Fig.~\ref{fig:wavepacket_oscip}c and d).   As the barrier height increases, wavepackets are more likely to be reflected.  In this case, the box modes are on resonance, as these wavepackets receive more energy when they impinge on an oppositely moving barrier (Fig.~\ref{fig:wavepacket_oscip}a and b).  This crossover from box-mode resonant behavior to ring-mode resonant behavior appears to occur near $V_0/\mu\approx0.3$, as seen in Fig.~\ref{fig:loss_spec_position}.  We note that in addition to the discrepancy in $\nu$, the simulated spectra show more atom loss than the experiment. 

The simulated spectrum with $V_0/\mu=0.6$ also shows some possible broadening, as seen by the additional atom loss at nonresonant frequencies $\nu_d\approx47$~Hz and 85~Hz.   As atoms are lost from the condensate, the speed of sound and $\nu$ decrease, causing the broadening.  As such, the broadening becomes evident only after $t\approx 1$~s of oscillation.  Broadening in the opposite direction (toward larger $\nu$) could also be present.  In particular, driving the condensate with sufficient strength can generate supersonic shock waves rather than sound waves.  In the next section, we directly create such dispersive shock waves and study their behavior.  

\section{\label{sec:shock_waves} Generation of supersonic shock waves}

\begin{figure}
\centering
\includegraphics[width=6.in]{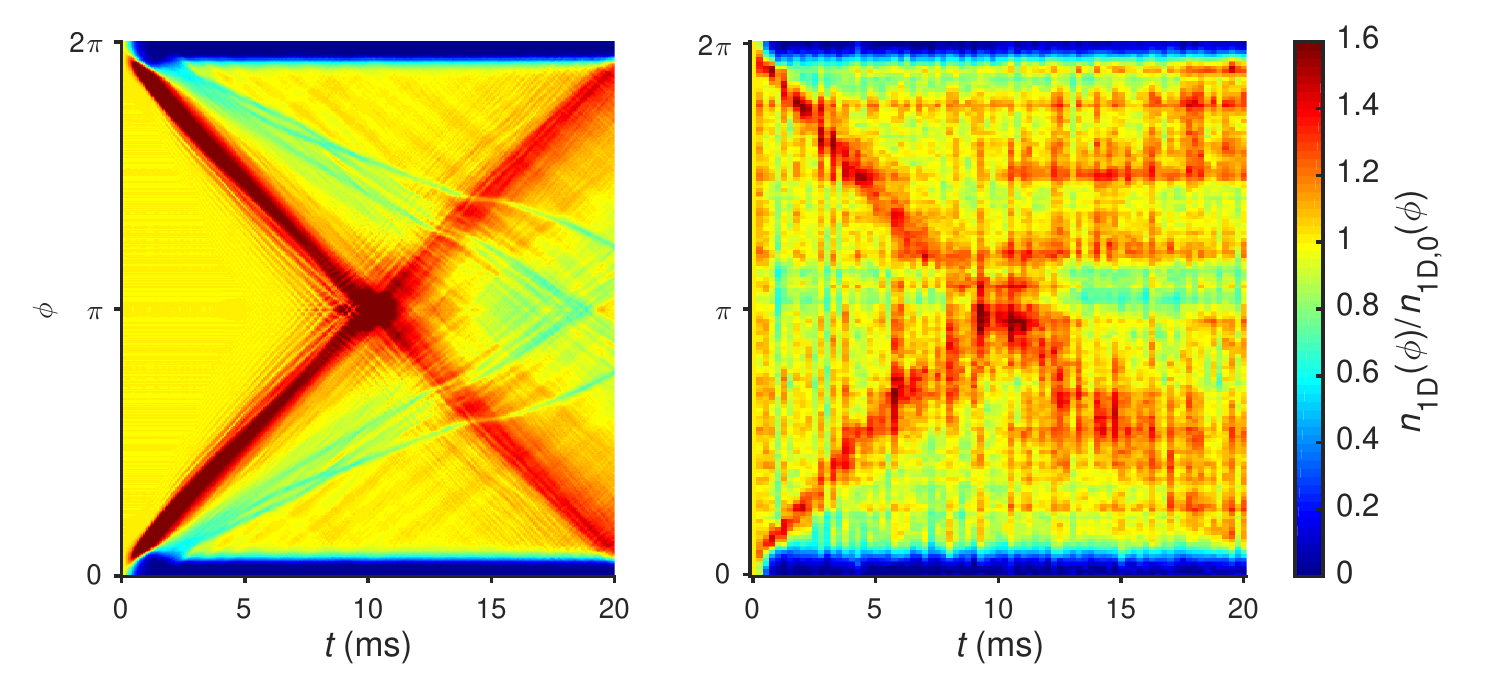}
\caption{\label{fig:el_shocko} Supersonic flow in a shocked BEC. A condensate in our standard configuration is struck with an amplitude pulse that rises to its peak strength during an interval of 100 $\mu\mathrm{s}$.  Normalized density is shown vs. time. Left: solution of the time-dependent GPE; right: experiment.}
\end{figure}

Theoretical analyses of the GPE predicted the existence of supersonic shock waves in BECs subject to large-amplitude disturbances~\cite{2003Zak_Kulikov,2003Kulikov_Zak,2004Damski}.  Observations of such waves in $^{87}$Rb condensates were later reported in Refs.~\cite{PhysRevLett.101.170404} and~\cite{PhysRevA.80.043606}.  There is a substantial theoretical literature on supersonic phenomena 
in BECs (see refs.~\cite{PhysRevA.75.033619,PhysRevLett.100.160402,bethuel2009travelling,Frantzeskakis2010,
PhysRevA.85.033603,2012NatPhysMathy,2012Zapata,PhysRevA.91.053603,DNSE2015} and references therein), and a recent experiment reports the experimental observation of analogue Hawking radiation in a BEC~\cite{nphys3104}.
We have also found evidence for supersonic shock waves, and our ring geometry makes it possible to observe collisions between them.

The left frame of Fig.~\ref{fig:el_shocko} shows the solution of the TDGPE for our standard BEC configuration, subject to a sudden raising of the barrier during 100 $\mu\mathrm{s}$.
(Because there is no atom loss, we do not rescale the theoretical simulation in Fig. 8 as we did in Figs. 4 and 5.) This results in two counterpropagating high-density pulses - the left side of a red ``X'' -  with orbital frequencies of $\nu \approx 50$ Hz, which is about 25\% greater than the orbital frequency of sound. These pulses collide near $\phi = \pi$, resulting in secondary excitations. However, the original shock pulses maintain much of their mass after the collision and continue to travel at the same speed.  The shock waves generated in the experiment, which are shown in the right frame, exhibit significantly greater dispersion after the collision.  

Also visible in the left frame of Fig.~\ref{fig:el_shocko} are some secondary striations at densities $n \approx 1$, which all propagate with speeds corresponding to orbital frequencies $\nu \approx 34 $ Hz.  These features are consistent with sound waves.  There are other structures at densities around $n \approx 0.6$, with orbital frequencies $\nu \approx 29 $ Hz.  Note that their speeds decrease during collisions with the shock waves, but are restored after the collision, a behavior characteristic of gray solitons.  Both sound waves and solitons were reported in early experiments on large-amplitude excitations of Na and Rb condensates \cite{Burger99,Denschlag2000,Dutton2001}.  However, we do not see definitive signatures of them in the experimental data, which is shown in the right frame of Fig. \ref{fig:el_shocko}.

\section{Summary \label{sec:summary}}

We have investigated the excitations of a ring-shaped condensate with oscillatory amplitude- and position-modulated perturbations.  This perturbation, in the form of a weak link, generates phonon wavepackets that travel around the ring at the speed of sound and therefore have an orbital angular frequency $\nu = c/(2 \pi R)$.  We find that the wavepackets are resonant with an amplitude-modulated perturbation if the perturbation's frequency is an integer multiple of $\nu$.  For position-modulation, the wavepackets are in resonance if the frequency of the perturbation is a half-integer multiple of $\nu$.  The difference in these cases corresponds to the symmetry of the drive: an amplitude modulation creates two oppositely moving wavepackets at the same time, whereas  position modulation creates two oppositely moving wavepackets at points in its motion that are out of phase by $\pi$.  By looking at atom loss as a function of drive frequency, we verify these resonance conditions.  
%By changing the height of the barrier to incoming phonon wavepackets can change these resonance conditions. {\bf REVIEW BEFORE EXIT}

This work has implications for other atomtronic devices.  For example, one should be able to induce a Shapiro resonance~\cite{Shapiro1963,Sols2004} in ultracold atoms by driving a weak link perturbation in a way similar to that done here.  In addition, phonon modes can be excited and controlled for future applications, such as phonon interferometry~\cite{Marti2015} and the detection of circulation states of a ring~\cite{Kumar2015}.   In the strongly reflecting regime, phonon wavepackets undergo similar time evolution as particles in a shaken box~\cite{Makowski1991,Drossel1998}, and thus could be useful for future studies of quantum chaos and Fermi acceleration~\cite{Seba1990,Grubelnik2014}.

\bibliography{ring}

\end{document}